\documentclass{aa}  
\usepackage{graphicx}
\usepackage{natbib}
\bibpunct{(}{)}{:}{a}{}{,}
\usepackage{txfonts}
\begin{document}
   \title{Precise radial velocities of giant stars. IV.\thanks{Based on data
   obtained at UCO/Lick Observatory, USA}}

\subtitle{A correlation between surface gravity and radial velocity variation and a statistical investigation of companion properties.}

   \author{S. Hekker \inst{1}\fnmsep \inst{2}
    \and I.A.G. Snellen \inst{1}
     \and C. Aerts \inst{3} \fnmsep \inst{4}
       \and A. Quirrenbach \inst{1} \fnmsep \inst{5}
        \and S. Reffert \inst{5}
           \and D.S. Mitchell\inst{6}}

   \offprints{S. Hekker, \\
                    email: saskia@oma.be}
   
   \institute{Leiden Observatory, Leiden University, P.O. Box 9513, 2300 RA
Leiden, The Netherlands \and Royal Observatory of Belgium, Ringlaan 3, 1180 Brussels, Belgium \and Instituut voor Sterrenkunde, Katholieke
Universiteit Leuven, Celestijnenlaan 200 D, 3001 Leuven, Belgium \and Department of
Astrophysics, University of Nijmegen, P.O. Box 9010, 6500 GL Nijmegen, The
Netherlands \and ZAH, Landessternwarte Heidelberg, K\"{o}nigstuhl 12, D-69117
Heidelberg, Germany \and California Polytechnic State University, San Luis
Obispo, CA 93407, USA }

   \date{Received <date>; accepted <date>}

 
  \abstract 
{Since 1999, we have been conducting a radial velocity survey of 179 K giants using 
the Coud\'e Auxiliary Telescope at UCO/Lick observatory. At present 
$\sim$20$-$100 measurements have been collected per star with a precision of 5 to 8 m s$^{-1}$. Of the stars monitored, 145 (80\%) show radial 
velocity (RV) variations at a level $>$20 m s$^{-1}$, of which 43 exhibit 
significant periodicities.}
{Our aim is to investigate possible mechanism(s) that cause 
these observed RV variations. We intend to test whether
these variations are intrinsic in nature, or possibly induced by companions, or both. In addition, we aim to characterise the parameters of these companions.}
{A relation between $\log g$ and the amplitude of the RV variations 
is investigated for all stars in the sample. Furthermore, the hypothesis that all periodic RV variations
are caused by companions is investigated by comparing their inferred orbital
statistics with the statistics of companions around main sequence F, G, and K
dwarfs.}
{A strong relation is found between the amplitude of the RV variations
and  $\log g$ in K giant stars, as suggested earlier by 
Hatzes \& Cochran (1998). However, most of the stars exhibiting periodic 
variations are located above this relation. These RV variations can be split in a periodic component which is not correlated with $\log g$ and a random residual part which \textsl{does} correlate with $\log g$. Compared to main-sequence dwarf stars, K giants frequently exhibit periodic RV variations. Interpreting these RV variations as being caused by companions, the orbital parameters are different from the companions orbiting dwarfs.}
{Intrinsic mechanisms play an important role in producing RV variations in 
K giants stars, as suggested by their dependence on $\log g$. However, it appears that 
periodic RV variations are \textsl{additional} to these intrinsic variations, 
consistent with them being caused by companions. \textbf{\rm If indeed the majority of the periodic RV variations in K giants is interpreted as due to substellar companions, then massive
planets are significantly more common around K giants than around F, G, K
main-sequence stars.}}

\keywords{stars: variables -- techniques: radial velocities} 
\authorrunning{S. Hekker et al.}  
\maketitle
%
\section{Introduction}
For more than a decade, radial velocity observations with accuracies of order
m\,s$^{-1}$ have been within reach (see for instance \citet{marbut2000} and
\citet{queloz2001}). Even accuracies of less than 1 m\,s$^{-1}$ \citep{pepe2003}
are possible now. With these observations, more than 200 sub-stellar
companions have been discovered by measuring the reflex motions of their parent stars. Most
of these sub-stellar companions have been detected around F, G and K main
sequence stars, but detections around an A star \citep{galland2006} and several
subgiants (\citet{johnson2006}, \citet{johnson2007}) have also been reported
recently.  Moreover, 10 giant stars were reported to have sub-stellar companions
($\iota$ Draconis (K2III) \citet{frink2002}, HD104985 (G9III) \citet{sato2003},
HD47526 (K1III) \citet{setiawan2003}, HD13189 (K2II-III) \citet{hatzes2005},
HD11977 (G5III) \citet{setiawan2005}, Pollux (K0III) \citet{hatzes2006},
\citet{reffert2006}, 4UMa (K1III) \citet{dollinger2007}, NGC2423
No3 and NGC4349 No127 \citet{lovis2007}, and recently HD17092 (K0III) \citet{niedzielski2007})\footnote{For updated information on
sub-stellar companions, see http://exoplanet.eu and http://exoplanets.org.}. In
addition to searches for extra-solar companions, radial velocity observations
prove to be very useful for detecting solar-like oscillations in stars with
turbulent atmospheres, such as the dwarf $\alpha$ Cen A
\citep[e.g.][]{bedding2006}, the subgiant Procyon \citep[e.g.][]{eggenberger2004,martic2004} and the giant $\epsilon$ Ophiuchi \citep[e.g.][]{deridder2006}.

With techniques for accurate radial velocity observations at hand, a survey was
started in 1999 to verify whether K giants are stable enough to be used as
astrometric reference stars for SIM/PlanetQuest (Space Interferometry Mission)
\citep{frink2001}. This survey contains 179 stars and uses the Coud\'e Auxiliary
Telescope (CAT) at University of California Observatories / Lick Observatory, in
conjunction with the Hamilton Echelle Spectrograph. The survey has recently been
expanded to about 380 giants and is still ongoing. For the analysis
described in the present paper only data from the initial 179 stars are used.

From this survey, companions have been announced for $\iota$ Draconis
\citep{frink2002} and Pollux \citep{reffert2006}. Stars with radial velocity
variations of less than 20 m\,s$^{-1}$ have been presented as stable stars by
\citet{hekker2006a}. In addition, some binaries discovered with this survey, as
well as an extensive overview of the sample, will be presented in forthcoming
papers. 

As almost all of the stars show significant radial velocity variations,
we investigate here which mechanism causes these variations. Non-periodic radial
velocity variations, of the order of the investigated timescales, are most likely caused by some intrinsic mechanism, while the periodic variability can also be caused by companions. We also investigate the characteristics of these companions.

In Sect. 2, the radial velocity observations are described in detail. In Sect. 3, the relation
between the observed radial velocity amplitude and surface gravity is
investigated. In Sect. 4, we explore the hypothesis that all periodic radial velocity variations are caused by
sub-stellar companions, and we compare the inferred orbital parameters with those
obtained for sub-stellar companions orbiting main sequence stars. Our
conclusions are presented in Sect. 5.

\begin{figure}
\begin{minipage}{\linewidth}
\begin{minipage}{\linewidth}
\centering
\includegraphics[width=\linewidth]{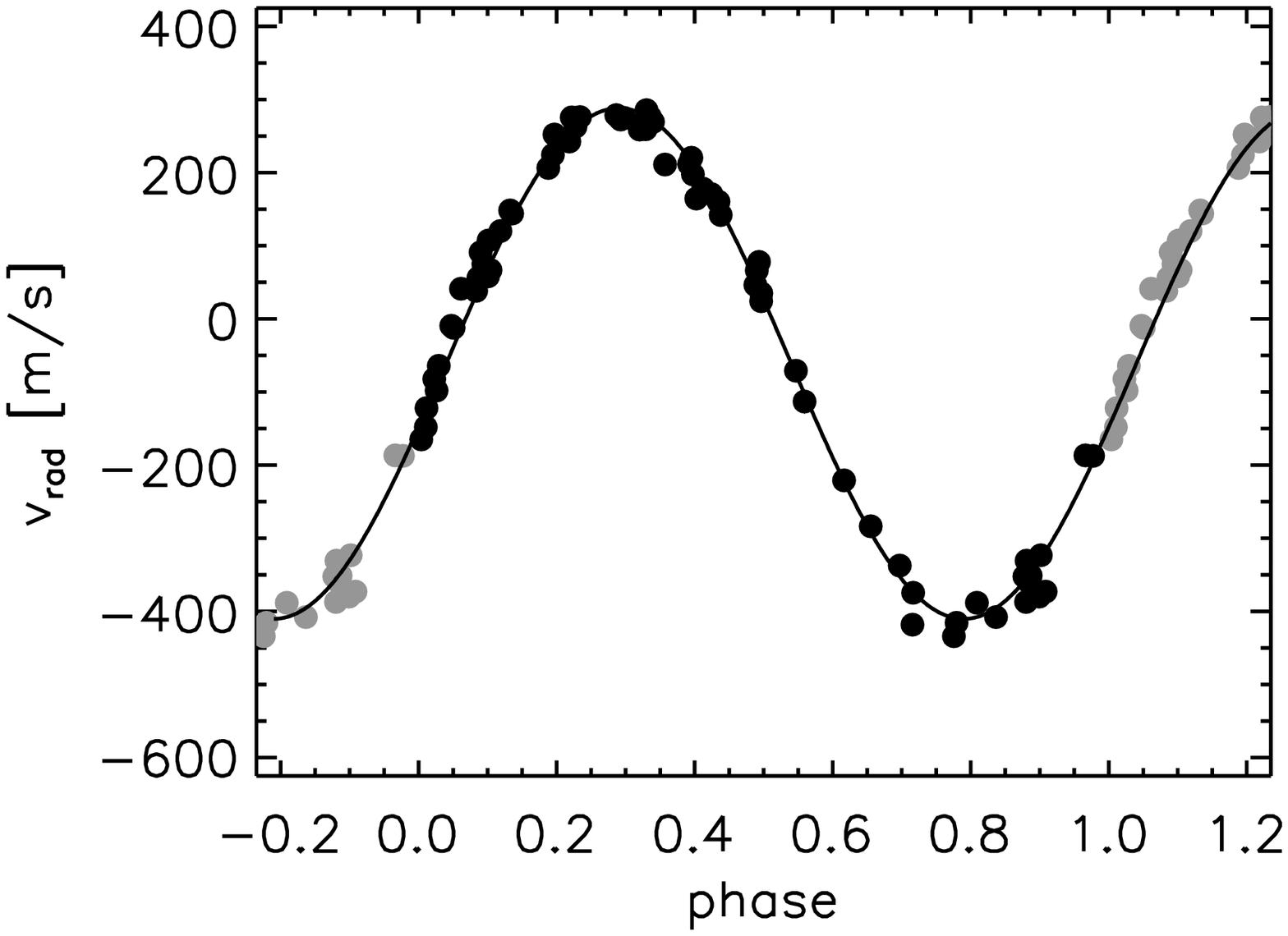}
\end{minipage}
\begin{minipage}{\linewidth}
\centering
\includegraphics[width=\linewidth]{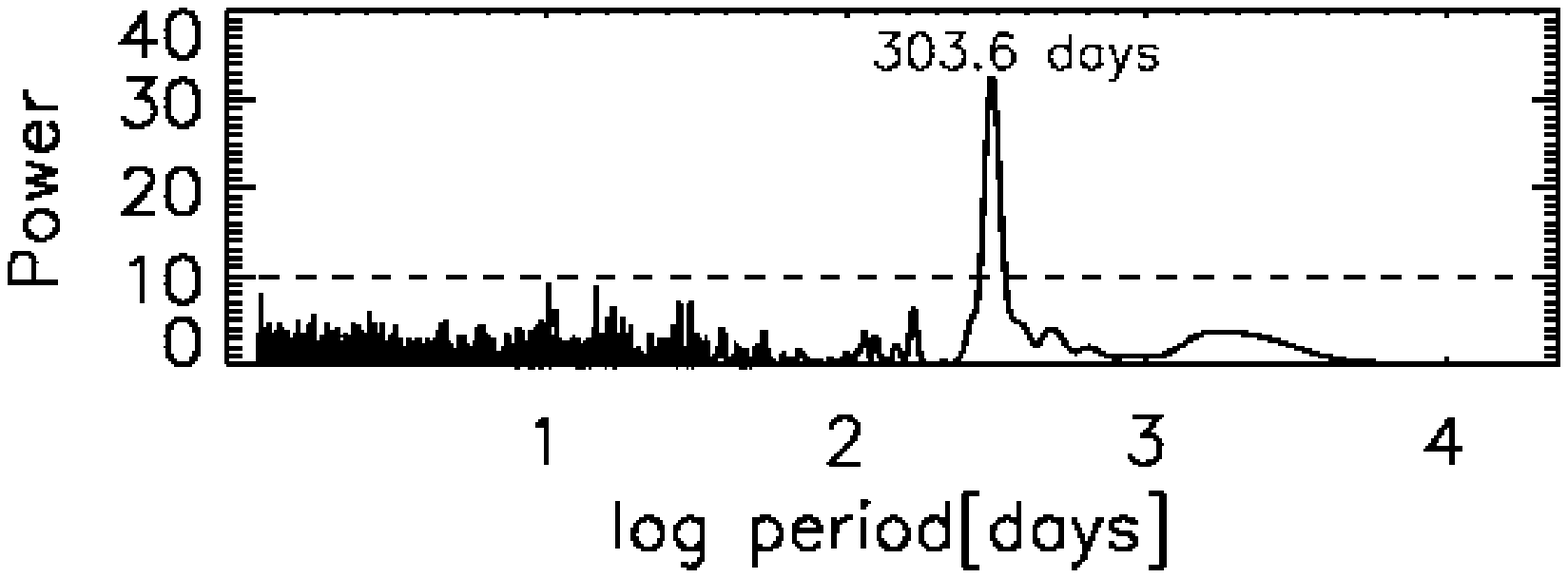}
\end{minipage}
\end{minipage}
\caption{Radial velocity variations as a function of phase for a star (HIP34693) with a
highly significant period (top), with its periodogram (bottom). The dashed line in the periodogram indicates the significance threshold.}
\label{vradphase1}
\end{figure}

\begin{figure}
\begin{minipage}{\linewidth}
\begin{minipage}{\linewidth}
\centering
\includegraphics[width=\linewidth]{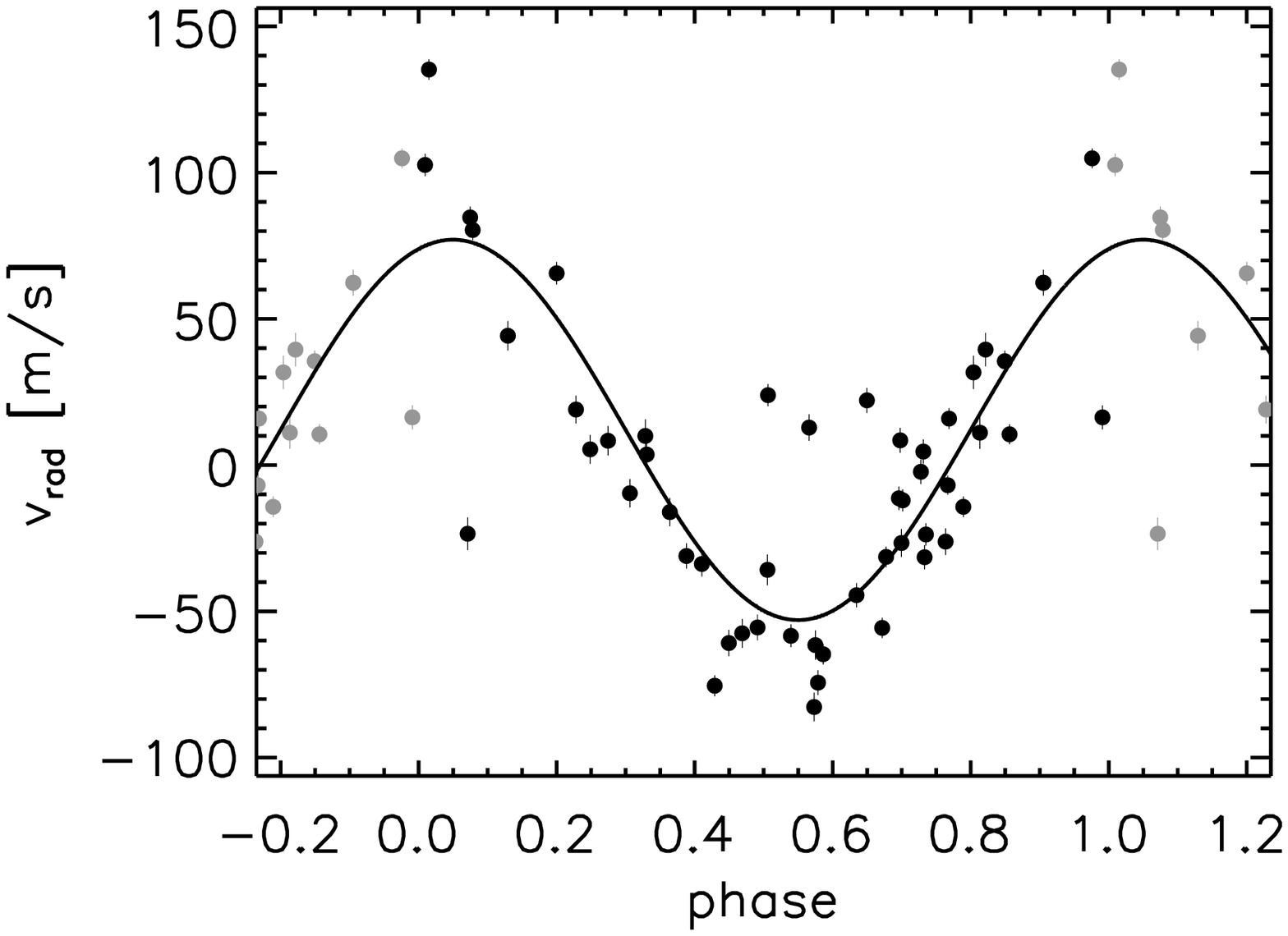}
\end{minipage}
\begin{minipage}{\linewidth}
\centering
\includegraphics[width=\linewidth]{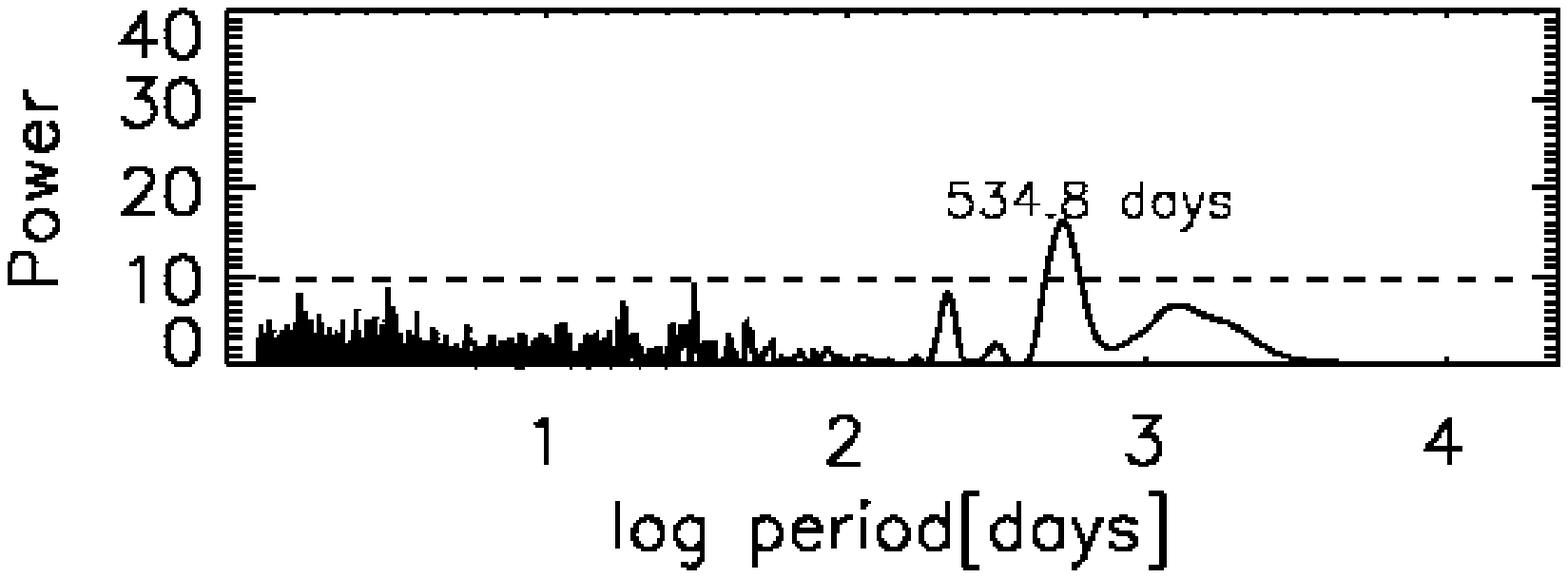}
\end{minipage}
\end{minipage}
\caption{Radial velocity variations as a function of phase for a star (HIP7607) with a period close to the significance threshold (top), with its periodogram
(bottom). The dashed line in the periodogram indicates the significance threshold.}
\label{vradphase2}
\end{figure}

\section{Radial velocity variations}

The initial 179 stars selected for the radial velocity survey are
used in the present work. These stars have been selected from the Hipparcos
catalogue \citep{esa1997}, based on the criteria described by
\citet{frink2001}. The selected stars are all brighter than 6~mag, are
presumably single and have photometric variations $< 0.06$~mag in V.  The survey
started in 1999 at Lick Observatory using the Coud\'e Auxiliary Telescope (CAT) in
conjunction with the Hamilton Echelle Spectrograph (R=60\,000). The system with
an iodine cell in the light path has been described by
\citet{marcy1992} and \citet{valenti1995}. With integration times of up to
thirty minutes for the faintest stars ($m_{v}$ = 6 mag) we reach a signal to
noise ratio of about $80-100$ at $\lambda = 5500$ \AA , yielding a radial
velocity precision of $5-8$ m\,s$^{-1}$. As we are looking for radial velocity
variations of order 10 to 100 m\,s$^{-1}$, this is adequate and hence no attempt
has been made to reach the 3 m\,s$^{-1}$ accuracy which is in principle possible
with this setup \citep{butler1996}. For the determination of the radial
velocities the pipeline described by \citet{butler1996} is used. In this
pipeline, a template iodine spectrum and a template spectrum of the target star
obtained without an iodine cell in the lightpath are used to model the stellar
observations with a superposed iodine spectrum. The Doppler shift is a free
parameter in this model. Note that with this method the absolute radial
velocity is not measured, but the radial velocity relative to the stellar
template is obtained.

All 179 stars are subjected to a period search.  The periodicity of the
radial velocity variations is determined first of all from a classical
Lomb-Scargle (LS) periodogram \citep{scargle1982}.  The significance threshold
is set to 6$\sigma$, where the noise level is determined from the average power
of the residual Scargle periodogram for frequencies between 0 and 0.03 cycles
per day (c/d) (0.35 $\mu$Hz) and a frequency step of 0.00001 c/d (0.12 $\cdot$
10$^{-3}$ $\mu$Hz). We adopted the conventional method of iterative sinewave fitting (`prewhitening') to
search for subsequent frequencies \citep{kuschnig1997}.  In Figs~\ref{vradphase1} and
\ref{vradphase2}, the radial velocity variation as a function of phase is shown
for two stars. The period of the star in Fig.~\ref{vradphase1} is highly
significant, while the one in Fig.~\ref{vradphase2} is close to the
significance threshold. Periodograms are shown in the bottom panels of these
figures.  As properly emphasized by \citet{cumming1999}, such a classical period
search may not be appropriate for unevenly spaced sparse data, even though we
set the significance level at a conservatively high level. In order to check
this, we have done an additional LS analysis after prewhitening the original
data by a linear polynomial. This led almost always to the same frequencies.  We
only accepted a frequency when it was found to meet the significance criterion for
both these analyses. The significant frequencies are listed in Table~\ref{freqs}. 

\begin{table}
\begin{minipage}[t]{\linewidth}
\caption{Single stars with significant frequencies.}
\label{freqs}
\centering
\begin{tabular}{lrrrr}
\hline\hline
star  & frequency & period & frequency & period\\
 & $\mu$Hz & days & $\mu$Hz & days\\
\hline
HIP3419 & 0.0638 & 181 & 0.1197 & 97\\
HIP7607 & 0.0216 & 536 & &\\
HIP7884 & 0.0184 & 629 & &\\
HIP13905 & 0.0223 & 519 & &\\
HIP16335 & 0.0197 & 588 & &\\
HIP19011 & 0.0252 & 459 & &\\
HIP21421 & 0.0203 & 570 & 0.0448 & 237\\
HIP23015 & 0.0151 & 767 & 0.0073 & 1586\\
HIP23123 & 0.0135 & 857 & &\\
HIP31592 & 0.0144 & 804 & &\\
HIP33160 & 0.0226 & 512 & &\\
HIP34693 & 0.0380 & 305 & &\\
HIP36616 & 0.0033 & 3507 & 0.0389 & 298\\
HIP37826 & 0.0194 & 597 & &\\
HIP38253 & 0.0171 & 677 & &\\
HIP39177 & 0.0133 & 870 & &\\
HIP40526 & 0.0172 & 673 & &\\
HIP46390 & 0.0233 & 497 & &\\
HIP47959 & 0.0187 & 619 & &\\
HIP53229 & 0.0022 & 5261 & &\\
HIP53261 & 0.0149 & 777 & &\\
HIP57399 & 0.0196 & 591 & &\\
HIP64823 & 0.0021& 5512 & &\\
HIP69673 & 0.0382 & 303 & 0.0187 & 619\\
HIP73133 & 0.0014 & 8267 & 0.0157 & 737\\
HIP73620 & 0.0226 & 512 & &\\
HIP74732 & 0.0233 & 497 & &\\
HIP75458 & 0.0226 & 512 & &\\
HIP79540 & 0.0203 & 570 & &\\
HIP80693 & 0.0022 & 5261 & 0.0198 & 585 \\
HIP84671 & 0.0251 & 461 & &\\
HIP85139 & 0.2895 & 40 & &\\
HIP85355 & 0.0258 & 449 & 0.0056 & 2067\\
HIP85693 & 0.0175 & 661 & &\\
HIP87808 & 0.0153 & 757 & &\\
HIP88048 & 0.0218 & 531 & 0.0029 & 3991\\
HIP91117 & 0.0303 & 382 & &\\
HIP109023 & 0.0199 & 582 & &\\
HIP109492 & 0.0217 & 533 & &\\
HIP109602 & 0.0130 & 890 & &\\
HIP109754 & 0.0189 & 612 & &\\
HIP113562 & 0.0401 & 289 & 0.0472 & 245\\
HIP114855 & 0.0639 & 181 & &\\
\hline
\end{tabular}
\end{minipage}
\end{table}

\section{Radial velocity amplitude - surface gravity relation}
\begin{figure*}
\centering
\includegraphics[width=\linewidth]{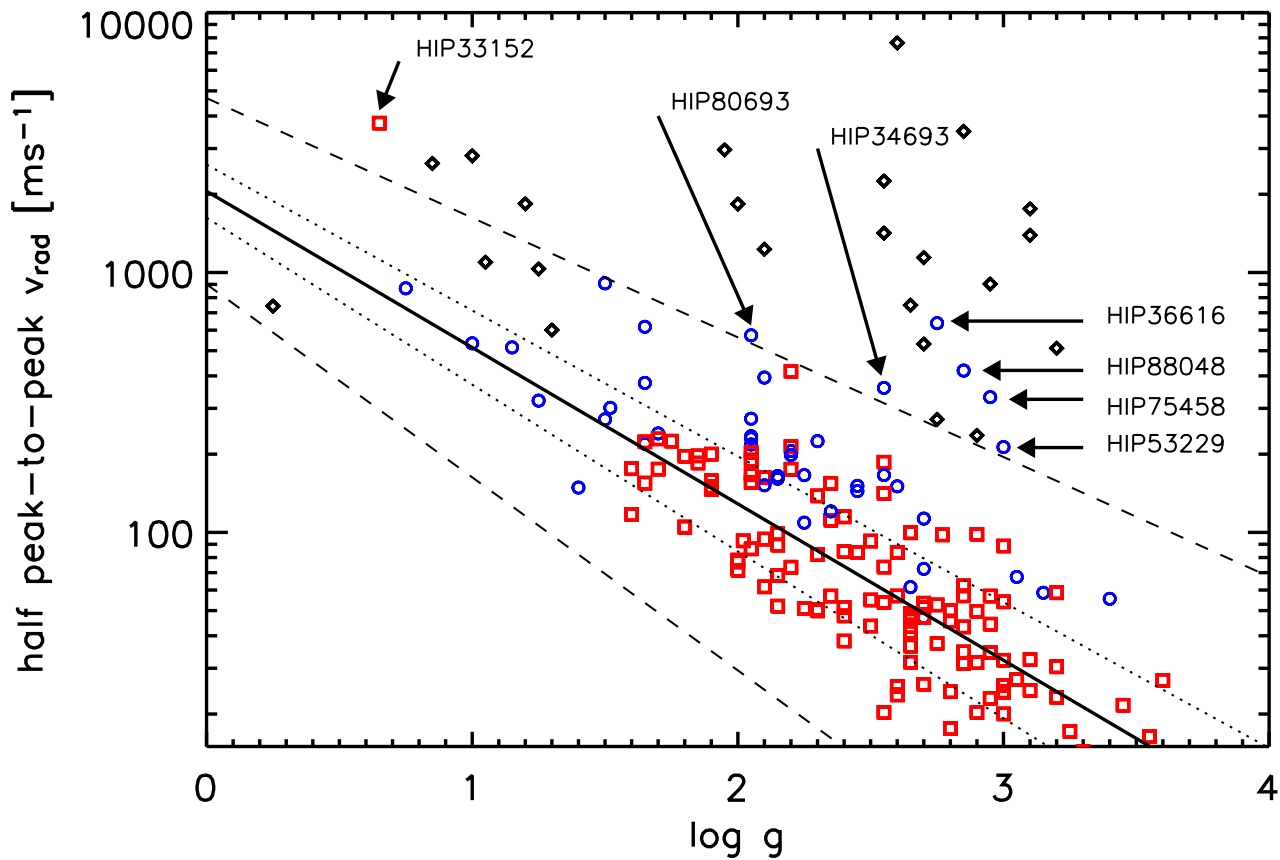}
\caption{Half of the peak-to-peak variation of the radial velocity as a function
of surface gravity ($\log g$). The blue $\circ$ indicate the single stars
with periodic radial velocity variations (see text for periodicity criteria),
the red $\square$ indicate single stars with random radial velocity variations, and
stellar binaries (companion mass $> 100$ M$_{\rm Jup}$) are indicated with black
$\diamond$ symbols. (Colours are only visible in the online version of the paper.) The solid line is the
best fit through the random single stars, the dotted line indicates the $1\sigma$
interval around the best fit and the dashed line indicates the $3.5\sigma$
interval. Single stars with a higher radial velocity amplitude than expected
based on their $\log g$ value (more than $3.5\sigma$ above the best fit) are
indicated by arrows. Six of the 8 stars with periodic radial velocity variations and $\log g < 1.6$ are classified bright giants or supergiants \citep{esa1997}.}
\label{klogg}
\end{figure*}

\citet{hatzes1998} already investigated the origin of the observed radial
velocities in K giant stars. Although their sample contained only 9 stars, they
suggested that the amplitude of the radial velocity increases with decreasing
surface gravity ($\log g$).  In lower surface gravity it takes longer to
decrease the velocity of a moving parcel which results in larger amplitudes and
the relation suggested by \citet{hatzes1998} would therefore be evidence for
pulsations or rotational modulation as the mechanism for these long period
radial velocity variations.

For the present sample, $\log g$ values were determined spectroscopically by \citet{hekker2007}, by imposing excitation and ionisation
equilibrium of iron lines through stellar models. The equivalent width of about
two dozen carefully selected iron lines were used for a spectroscopic LTE
analysis based on the 2002 version of MOOG \citep{sneden1973} and Kurucz model
atmospheres which include overshooting \citep{castelli1997}. These authors
estimated the error on $\log g$ to be 0.22 dex from the scatter found in a
comparison with literature values. A detailed description of the stellar
parameters for individual stars and a comparison with literature values
are available in  \citet{hekker2007} and is therefore omitted here.

In Fig.~\ref{klogg}, we show half of the peak-to-peak value of the observed
radial velocity variations as a function of $\log g$ for K giants in our
sample. A clear trend is visible between increasing radial velocity variations
in single stars, and decreasing $\log g$, which provides a strong
indication that, at least for a large fraction of stars in our sample, the
observed radial velocity variations are induced by a mechanism intrinsic to the
star. This trend is present for stars with random as well as stars with
periodic radial velocity variations. Also, nearly all stars with periodic radial velocity variations and $\log g \geq 1.6$ are located above the fit in Fig.~\ref{klogg}. Seven single stars have a
higher radial velocity variation than expected based on their $\log g$ value,
i.e.\ they are situated more than $3.5\sigma$ above the best fit for the
relation obtained for single stars. These stars are indicated with arrows in
Fig.~\ref{klogg}. The radial velocity variation observed for HIP53229 may be
due to a stellar companion in a wide orbit, with a period much longer than the
observation time span. Due to this long period the companion mass, and,
therefore, the (sub-)stellar nature, is still very uncertain. HIP33152
is classified as a supergiant. The observed radial velocity variations for
HIP80693, HIP36616, and HIP88048 can be fitted with two Keplerian
orbits, while HIP75458 can be explained by an eccentric sub-stellar companion
\citep{frink2002} and an additional linear trend, indicating a companion in a
wide orbit. HIP34693 can be fitted very accurately with a Keplerian orbit of a
single nearly sinusoidal sub-stellar companion. 

\begin{figure*}
\centering
\includegraphics[width=\linewidth]{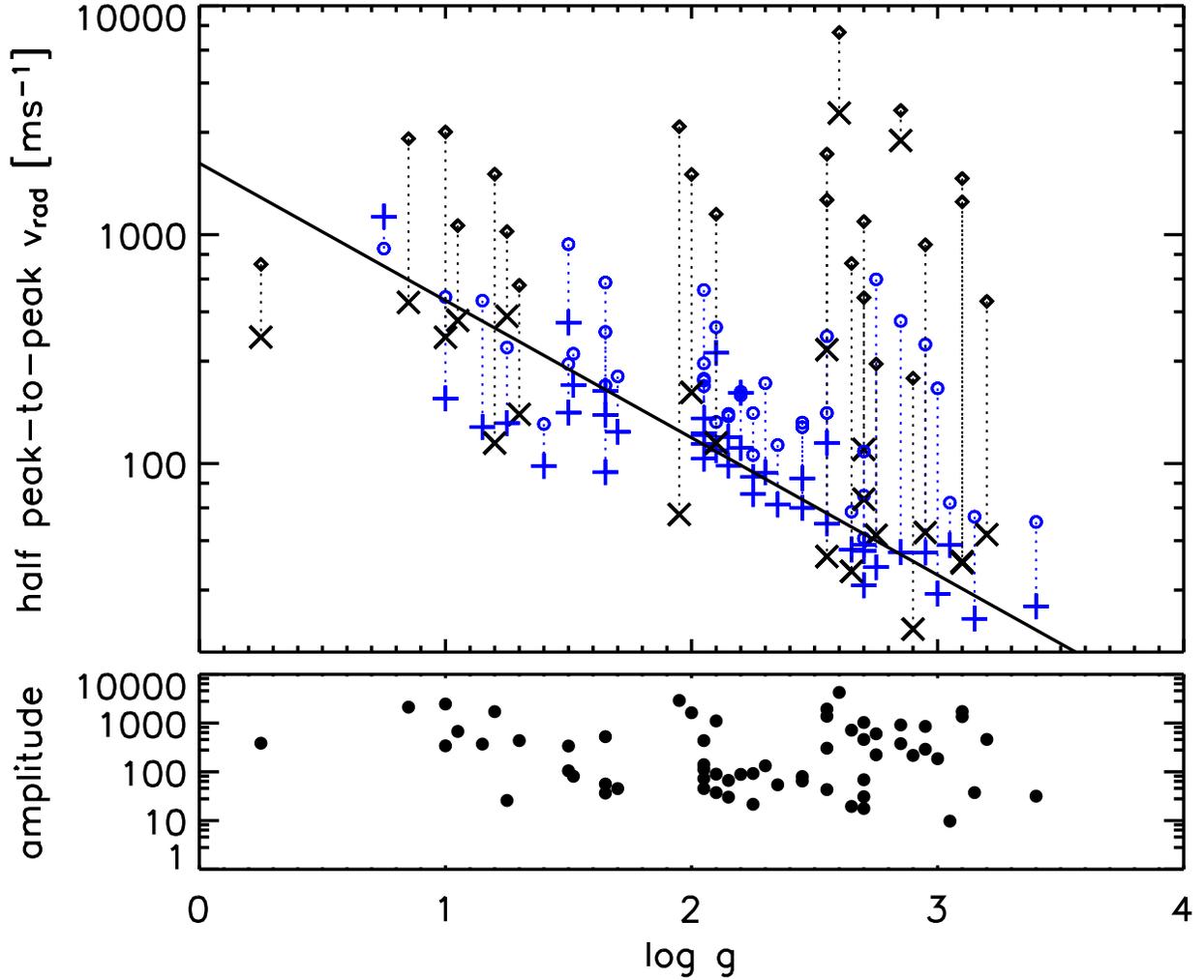}
\caption{Top: Half of the peak-to-peak variation of the radial velocity as a function
of surface gravity ($\log g$), as in Fig.~\ref{klogg}, but showing only those stars with periodic radial velocity variations (blue $\circ$) and stellar binaries (black $\diamond$). The blue $+$ and black $\times$ symbols indicate the amplitude of the radial velocity variations for these stars after subtraction of the Keplerian fits. (Colours are only visible in the online version of the paper.) The solid line indicates the linear fit through the stars with non-periodic radial velocity variations (from Fig.~\ref{klogg}). Bottom: Amplitudes of the subtracted Keplerian fits as a function of $\log g$.}
\label{kloggres}
\end{figure*}

In order to investigate the simultaneous occurrence of sub-stellar companions and oscillations in giants, the best Keplerian fits are subtracted from the observed radial velocity variations of both the binaries and the single stars with significant periodic radial velocity variations. The half peak-to-peak values of these residuals are plotted as a function of surface gravity in Fig.~\ref{kloggres}, also showing the linear fit through the non-periodic stars, and the amplitudes of the subtracted periodic signals. There are several interesting points to make about this graph. Firstly, almost all periodic stars show larger radial velocity variations than predicted by the relation found for non-periodic stars, but when the periodicities are removed, their residual radial velocity variations follow the same relation as found for the non-periodic stars. This could be interpreted as evidence for both intrinsic (non-periodic radial velocity variation) and extrinsic (periodic variations) mechanisms playing a role in these stars. Secondly, there is no correlation between the amplitude of the subtracted periodic signal and $\log g$, which provides additional evidence for the presence of companions. Thirdly, almost all (8 out of 9) stars with $\log g \leq 1.6$ exhibit periodic variations.  If these indeed have an extrinsic mechanism, it would mean that $\sim 90\%$ of these stars have sub-stellar companions. However, from an astrophysical point of view, stars with such low surface gravities are already very high on the giant branch or even on the asymptotic giant branch. At these low gravities, stars cannot be constant anymore, as the outer layers are so diluted that instabilities occur easily, either periodic or random. Stated differently, these stars are very close to the semi-regulars, which are on their way to become Mira variables. Hence, these periodic variations could well be intrinsic. 

\section{Companion Interpretation}
From the analysis of the correlation between radial velocity amplitude and surface
gravity we have evidence for the presence of both intrinsic variability and companions in at least a fraction of the K giant stars with periodic radial velocity variations. In order to study the characteristics of these companions, we take the hypothesis that the periodic radial velocity variations detected in
43 of our K giant stars, excluding binaries, are caused by sub-stellar companions. Under this
hypothesis, we investigate the statistical properties of the orbital parameters
of the sample and compare these with the statistical properties of companions
orbiting F, G and K dwarfs.

According to our analysis, 55 stars in the sample would have a single companion,
and 11 stars multiple companions. Twenty-three (22 single and 1 in a multiple
system) of these companions have $m\sin i$ larger than 100 M$_{\rm Jup}$ and
should be interpreted as stellar binaries.  By advancing the multiple
sub-stellar systems forward in time via a Runga-Kutta integration, we
investigated the stability of the systems, taking into account the mutual
interactions of the companions. With the orbital parameters that minimise $\chi^2$ taken at face value, we found that most of the inferred sub-stellar
multiple systems would be 'likely unstable', with a change in semi-major axis $>
1\%$ and $<10\%$, or 'unstable' with a change in semi-major axis $>10\%$ on a
time scale of 100 years due to companion-companion interaction.

However, the inferred stability depends on the starting epoch of the computations, as well
as on the orbital parameters, which might change with an increasing number of
observations. Furthermore, there is no guarantee that the obtained $\chi^{2}$
minimum is a global minimum. Therefore, stars with multiple inferred sub-stellar
companions that seem to be unstable, might also have stable solutions.  One
could also use the equations for dynamical stability described by
\citet{gladman1993} and \citet{marcy2001}. \citet{gladman1993} also notes that
the Hill stability criteria for companions in initially eccentric orbits may not
be met, but that the systems may still be found to be empirically quite stable
for a long period of time. In order to draw a firm conclusion on the stability
of a particular system, a more thorough investigation is needed, as well as data
with a longer time base, which is beyond the scope of this paper.

For all stars with periodic radial velocity variations, we checked the Hipparcos
\citep{esa1997} photometry. We checked periodograms for significant frequencies
close to the obtained radial velocity period, and plotted the photometric values
phased with the radial velocity period. None of the stars show photometric
variations related to the observed radial velocity variations.

The mass distribution of our K giant sample is not known very well. The stellar
masses are typically between 1 and 4 M$_{\sun}$. Hence most of their main
sequence progenitors should have been of A or F spectral class. The distribution
of orbital parameters of sub-stellar companions orbiting A and F main sequence
stars is still unknown. 
The core accretion model predicts more giant planets around more massive stars, so
that the distribution of orbital parameters of sub-stellar companions orbiting
F, G and K main-sequence stars probably cannot serve as a proxy. However, it
should be instructive to compare the two distributions.

Since the data presented here span $\sim 2500$ days, radial velocity variations
with longer periods are uncertain, and, therefore, not taken into
consideration. Companions with periods exceeding the observation time span are
also excluded from the F, G, and K main sequence star statistics.

\subsection{Mass distribution}
\begin{figure}
\centering
\includegraphics[width=\linewidth]{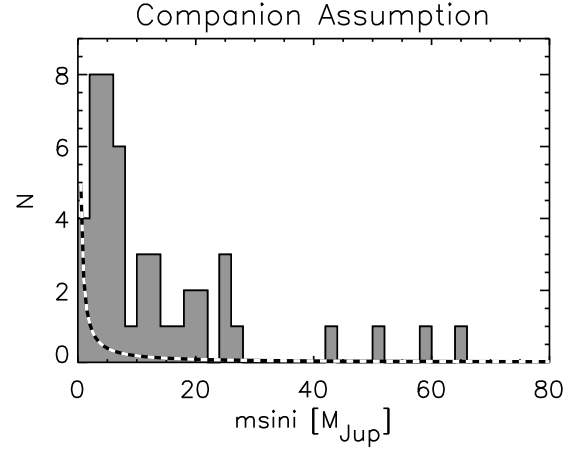}
\caption{A histogram of $m\sin i$ of inferred companion masses orbiting K giants
in our sample. The dashed line indicates the rise of planet masses $M^{-1.05}$
from 10 M$_{\rm Jup}$ down to Saturn masses for sub-stellar companions around
main sequence stars \citep{marcy2005}, normalised to the number of stars in our
sample.}
\label{msini}
\end{figure}

\begin{figure}
\centering
\includegraphics[width=\linewidth]{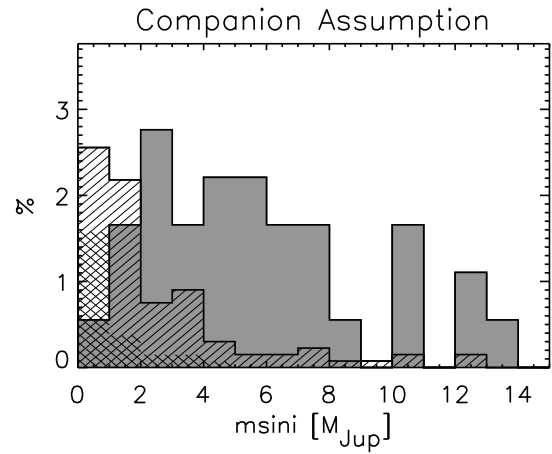}
\caption{Zoom in on the low end of the companion mass distribution of inferred
sub-stellar companions orbiting K giant stars in our survey, shown as percentage
of the total number of stars in the sample (gray histogram). The hatched
histogram shows the distribution of companion masses orbiting main sequence
stars as shown in Fig.~1 of \citet{marcy2005}, as a percentage of the total
number of stars in their sample. The cross hatched area are main sequence stars
with companions at a semi-major axis smaller than 0.3 AU (see
Fig.~\ref{semaxislog}).}
\label{msinilow}
\end{figure}

Fig.~\ref{msini} shows the distribution of inferred companion masses of our K
giant sample. First, notice that $30\%$ of the inferred companions would have
masses in the brown dwarf regime $15$ M$_{\rm Jup} < m\sin i< 80$ M$_{\rm
Jup}$. This is in sharp contrast to the brown dwarf statistics around F, G and K
main sequence stars for which only very few companions are found with $m\sin i >
15$ M$_{\rm Jup}$ around more than thousand stars. This is known as the brown
dwarf desert and is possibly caused by migration and merging of brown dwarfs in
a viscous disk with a mass at least comparable to the brown dwarf mass
\citep{armitage2002b}. 

The dashed line in Fig.~\ref{msini} indicates the rise of sub-stellar companion
masses $M^{-1.05}$ from 10 M$_{\rm Jup}$ down to Saturn masses for main sequence
stars \citep{marcy2005}, normalised to the number of stars in our sample.  We
use a two-sided Kolmogorov-Smirnov test \citep{press1992} (hereafter KS-test) to
compare the $M^{-1.05}$ fit and the mass distribution ($m\sin i< 80$ M$_{\rm
Jup}$) of our sample and find a probability of 0.05$\%$ (0.002$\%$ for $m \sin i
< 28$ M$_{\rm Jup}$) that these are identical. This implies that inferred
sub-stellar companions around K giants in our sample have higher masses compared
to companions around F, G and K main sequence stars.

Fig.~\ref{msinilow} shows the low-mass companion distribution of our survey.
Most K giant companions would have inferred masses between 2
and 8 M$_{\rm Jup}$, while the fraction of companions orbiting F, G and K dwarfs
strongly decreases with increasing $m\sin i$.

\subsection{Semi-major axis distribution}
\begin{figure}
\centering
\includegraphics[width=\linewidth]{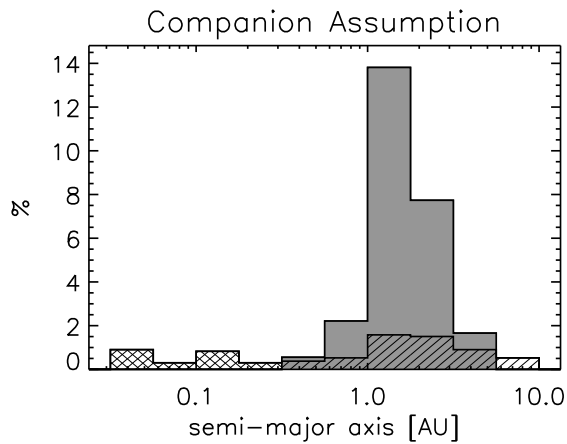}
\caption{Semi-major axis distribution of inferred K giant companions in our
sample (gray histogram), shown as a percentage of the total number of stars in
the sample. The hatched histogram shows the distribution for F, G and K main
sequence stars as shown in Fig.~2 of \citet{marcy2005}, as a percentage of the
total number of stars in their sample. The main sequence stars with a companion
orbiting at a semi-major axis smaller than 0.3 AU are cross-hatched. These are
also indicated in Figs~\ref{msinilow} and~\ref{period}.}
\label{semaxislog}
\end{figure}

The distribution of the companions' semi-major axis is shown in
Fig.~\ref{semaxislog}.  No inferred companions with semi-major axis smaller than
0.3 AU are present in the K giant sample, possibly due to increased stellar
radii of giants. The fraction of stars with an inferred companion with a
semi-major axis between 1 and 3 AU is much higher among the K giants compared to
the F, G and K dwarfs. A comparison between the two distributions with a KS-test
reveals a probability for the two distributions to be identical of 11$\%$. This
increases to 32$\%$ when omitting the main sequence stars with semi-major axis
$< 0.3$ AU. The increasing incompleteness beyond 3 AU, due to the limited time
span of surveys, is present in both samples as the surveys cover a comparable
amount of time. This incompleteness cannot cause the significant difference in
the peak between 1 and 3 AU.

\subsection{Period distribution}
\begin{figure}
\centering
\includegraphics[width=\linewidth]{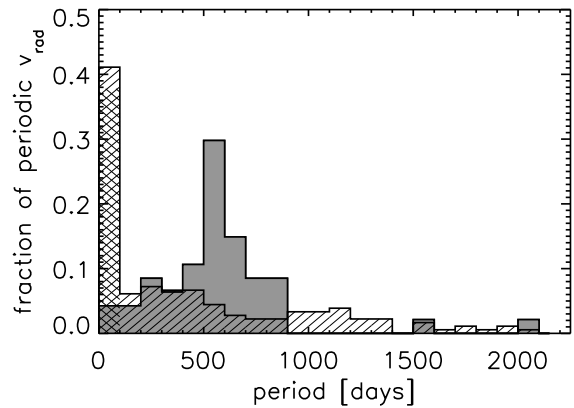}
\caption{Period distribution of the observed radial velocity variations shown as
a fraction of all significant periods (gray histogram). The period distribution
of main sequence stars \citep{butler2006} is shown in the hatched histogram, as
a fraction of the total number of observed companions. The cross hatched area
are main sequence stars with a semi-major axis smaller than 0.3 AU (see
Fig.~\ref{semaxislog}). }
\label{period}
\end{figure}

In Fig.~\ref{period} the period distribution of the observed radial velocity
variations is shown and compared with the companion period distribution of
dwarfs. The large fraction of F, G and K dwarf companions with orbital periods
shorter than 100 days corresponds to the ones with semi-major axis smaller than
approximately 0.3 AU. The close-in short-period companions are not present
around K giants, while about $80\%$ of these stars with observed radial
velocities have periods ranging between 400 and 800 days. A KS-test reveals a
probability of less than 0.0001$\%$ for the two distributions to be identical. The
probability remains below this level, when the companions orbiting main
sequence stars with a semi-major axis $< 0.3$ AU are omitted.

\subsection{Eccentricity distribution}
\begin{figure}
\centering
\includegraphics[width=\linewidth]{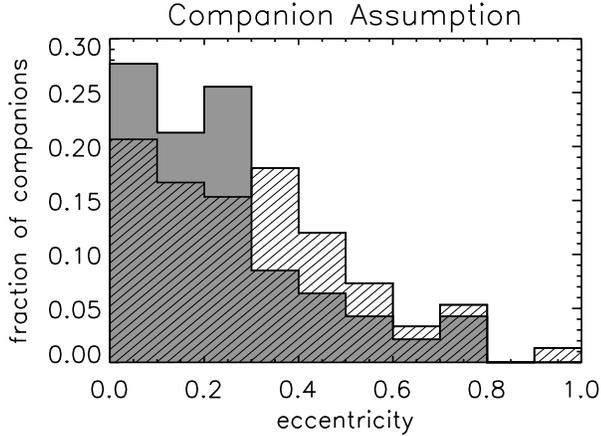}
\caption{Distribution of eccentricities for possible companions around K giants
in our sample (gray histogram) shown as a fraction of all possible sub-stellar
companions in the sample. The hatched histogram is the eccentricity distribution
of companions around main sequence stars \citep{butler2006} shown as a fraction
of all companions around main sequence stars. Companions with periods shorter
than 20 days are excluded from the latter sample as these might be tidally
circularised.}
\label{eccentricity}
\end{figure}

Fig.~\ref{eccentricity} shows the distribution of companion eccentricities for
K giants in our sample and for dwarfs. Companions of dwarfs with periods less
than 20 days are excluded, as these might be tidally circularised.
The fraction of companion eccentricities $<0.3$ for the giants is $75\%$
compared to $50\%$ for companions orbiting F, G and K dwarfs. The KS-test shows
that these distributions are nearly identical (97$\%$).

\subsection{Iron abundance}
\begin{figure}
\centering
\includegraphics[width=\linewidth]{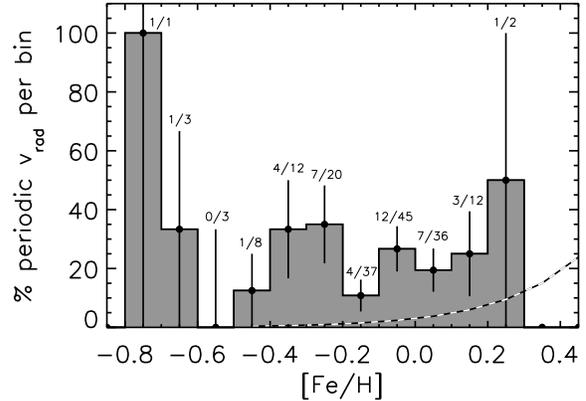}
\caption{Iron abundance [Fe/H] distribution of K giant stars with periodic
radial velocity variations shown as a percentage of the total number of observed
stars with iron abundance in the same interval. The numbers above each bar on
the histogram indicate the ratio of stars with a significant periodic radial
velocity variation to the total number of stars in each bin. The error bars are
calculated assuming Poisson statistics (i.e., the percentage of stars with
periodic radial velocities divided by the square root of the number of stars
with periodic radial velocities). The dashed line is the power law derived for
the increasing trend in the fraction of stars with companions as a function of
metallicity of F, G and K main sequence stars \citep{fischer2005}. }
\label{Fe}
\end{figure}

Companion occurrence correlates strongly with the abundance of heavy elements
(see for instance \citet{gonzalez1997}, \citet{fischer2005} and
\citet{santos2005}), such that F, G, and K dwarf stars with supersolar abundance
are more likely to harbour sub-stellar companions (about $50\%$ of the stars
with $0.3 < $ [Fe/H] $ < 0.5$). The increase of the fraction of F, G and K
dwarfs harbouring companions with increasing metallicity is well fitted with a
power law, yielding a probability for such a star to harbour a companion to be:
P = 0.03$\cdot$[($N_{\rm Fe}/N_{\rm H})/(N_{\rm Fe}/N_{\rm H})_{\sun}]^{2.0}$
\citep{fischer2005}.

In Fig.~\ref{Fe} the iron abundance distribution of stars with periodic radial
velocity variations is shown as a percentage of the total number of observed
stars with iron abundance in the same interval. The iron abundance is determined
spectroscopically by imposing excitation and ionisation equilibrium in iron
lines and is described by \citet{hekker2007}. The maximum iron abundance of a K
giant star in our sample is 0.29 and, therefore, we do not probe the high
metallicity region in which F, G and K dwarfs are most likely to harbour a
companion.

The mean metallicity of the K giants in the entire sample is $-0.12$ dex, while the mean metallicity of the stars with periodic radial velocity variations, presented in Fig.~\ref{Fe}, is $-0.13$ dex. No correlation between companion occurrence and abundance, similar to the one which is present for dwarf stars, is found in this sample of giant stars.

\section{Discussion and Conclusion}
The tight correlation we found between $\log g$ and half of the peak-to-peak
radial velocity variations seems to indicate that a large fraction of the
observed radial velocity variations in our sample of K giants is induced by
mechanism(s) intrinsic to the stars. We also present evidence that both intrinsic and extrinsic mechanisms play a role. The stars with a significant periodic signal are almost all located above the radial velocity amplitude vs. $\log g$ relation, but when the periodic signal is removed, the residuals show the same trend as for the non-periodic stars. Furthermore, no correlation is present between the amplitude of the periodic signal and $\log g$.

Almost all of the lowest $\log g$ stars show periodic variations. It may be possible that stars with such low surface gravity cannot be constant and that in these dilute atmospheres instabilities can occur very easily, and therefore may be periodic, but not extrinsic.

Based on the evidence that extrinsic mechanism(s) play a role for K giant stars with periodic radial velocity variations we investigated the hypothesis that this periodic signal is caused by the reflex motion of sub-stellar companions orbiting these stars. We presented the characteristics of the orbital parameters of these companions and compared them with the known orbital parameters of sub-stellar companions orbiting F, G and K dwarfs.

About 25$\%$ of the stars in our sample have radial velocity variations
with significant periodicity, and could possibly harbour a sub-stellar companion, while approximately only $8\%$ of the 1330 F, G and K main sequence stars investigated by \citet{marcy2005} have a sub-stellar companion. Recently \citet{johnson2007} and \citet{lovis2007} showed that the number of companion harbouring stars increases with mass. The giants in the present sample have typical masses between 1 and 4 M$_{\sun}$ and are in general more massive than the main sequence stars investigated for companions. So, the high percentage is qualitatively in agreement with the results from the literature. Furthermore, \citet{lovis2007} suggest that more massive stars form more massive planetary systems than lower mass stars. Figs~\ref{msini} and \ref{msinilow} show that we find, in general, more massive companions around the more massive K giants than are present around F, G and K dwarfs.

The high percentage of more massive companions around the more massive K giant stars would also be compatible with the core accretion model. This model predicts very few giant planets, but
a relatively large number of planets with the mass of Neptune or smaller around
M dwarfs \citep{laughlin2004,ida2005}. This is mainly the result of a
much reduced surface density of the disk and the resulting shorter disk
evolution timescales compared to those for more massive stars, implying that
planet properties vary with the mass of their host stars. In particular, one
would expect more sub-stellar companions with higher masses in our giant sample, as is indeed
the case if we assume that the companion hypothesis is correct. 

The mean metallicity of companion-hosting K giants would be similar to the mean
metallicity of the total sample. This is in contrast with the correlation between companion occurrence and metallicity present in F, G and K dwarfs \citep[e.g.][]{fischer2005}. So far, several groups have investigated the correlation between companion occurrence and metallicity for giants with different results. \citet{sadakane2005} and \citet{pasquini2007} agree that companion-ahosting giants are on average not metal-rich, while \citet{hekker2007} find that giants with announced companions have higher metallicities than their total sample of giants. A detailed discussion about these different results is presented by \citet{hekker2007}. They conclude that the samples on which the results are based are slightly different and the more metal-rich stars used in their study are lacking in the study by \citet{pasquini2007}. Furthermore, there is a difference in zero-point correction for the metallicities of announced companion-hosting stars from different surveys. All in all, these inferences are based on small-number statistics and all results have to be taken with caution.

The larger semi-major axis and long periods of the inferred companions orbiting K giant stars compared to companions orbiting dwarf stars are most likely due to the extended atmospheres of K giants. For the eccentricity no significant difference is found in the distribution between companions orbiting dwarfs or giants. Nevertheless, the high number of inferred companions around giants with eccentricities $<$ 0.3 is striking. One could suspect that companion orbits circularise over time and that the companions in circular orbits are older than the eccentric ones, but there is no evidence for this hypothesis.

In principle, nearly sinusoidal radial velocity variations could also be caused by pulsations or spots. Although the periods of the radial velocity variations could well be the rotational periods of the stars, the presence of prominent spots is not very likely. In that case one would also expect photometric variations with periods correlated with the radial velocity variations. From the Hipparcos photometry \citep{esa1997} such correlations were not found.

In order to distinguish with certainty between companions and pulsations as the cause of the observed radial velocity variations, one needs to perform a spectral line profile analysis. A technique for doing this with very high-resolution spectra ($R \geq$ 100\,000) will be presented separately (Hekker et al., in preparation) because the amount of such data at hand today is insufficient to add significantly to the conclusions of this paper.

\begin{acknowledgement}
We thank Debra Fischer, Geoff Marcy and Paul Butler for useful discussions and
the development of the instrumentation and software for the determination of the
radial velocities at Lick Observatory.  In addition, we thank the entire staff
at Lick Observatory for their excellent support. Finally, we would like to thank the anonymous referee for valuable comments and suggestions.
\end{acknowledgement}

\bibliographystyle{aa}
\bibliography{8321bib}
\end{document}